# Ground-based calibration and characterization of GRD of GECAM: 8-160 keV

J. J. He[1*], Z. H. An[1*], W. X. Peng[1*], X.Q. Li[1*], S. L. Xiong[1*], D. L. Zhang[1], R. Qiao[1], D. Y. Guo[1], C. Cai[1], Z. Chang[1], C. Chen[1], G. Chen[1], Y. Y. Du[1], M. Gao[1], R. Gao[1], K. Gong[1], D. J. Hou[1], C. Y. Li[1,6], G. Li[1], L. Li[1], M. S. Li[1], X. B. Li[1], X. F. Li[1], Y. G. Li[1], X. H. Liang[1], J. C. Liu[1], X. J. Liu[1], Y. Q. Liu[1], H. Lu[1], X. Ma[1], B. Meng[1], F. Shi[1], L. M. Song[1], X. L. Sun[1,2], C. W. Wang[1], H. Wang[1], H. Z. Wang[1], J. Z. Wang[1], Y. S. Wang[1], X. Wen[1], X. Y. Wen[1], S. Xiao[1], Y. B. Xu[1], Y. P. Xu[1], W. C. Xue[1], J. W. Yang[1], S. Yang[1], Q. B. Yi[1,3], C. M. Zhang[1], C. Y. Zhang[1], Fan Zhang[1], Fei Zhang[1], P. Zhang[1], S. N. Zhang[1,5], Y. Q. Zhang[1], X. Y. Zhao[1], Y. Zhao[1,4], C. Zheng[1], S. J. Zheng[1], X. Zhou[1], Y. Zhu[1]

[1]*Key Laboratory of Particle Astrophysics, Institute of High Energy Physics, Chinese Academy of Sciences, Beijing 100049, China*
[2]*State Key Laboratory of Particle Detection and Electronics, Beijing 100049, China*
[3]*Xiangtan University, Xiangtan 411105, China*
[4]*Beijing Normal University, Beijing 110875, China*
[5]*Space Science Division, National Astronomical Observatories of China, Chinese Academy of Sciences, Beijing 100012, China*
[6]*Physics and Space Science College, China West Normal University*

As the main detector of the GECAM satellite, the calibration of the energy response and detection efficiency of the GRD detector is the main content of the ground-based calibration. The calibration goal requires the calibrated energy points to sample the full energy range (8 keV-2 MeV) as much as possible. The low energy band (8-160 keV) is calibrated with the X-ray beam, while the high energy band (>160 keV) with radioactive sources. This article mainly focuses on the calibration of the energy response and detection efficiency in the 8-160 keV with a refined measurement around the absorption edges of the lanthanum bromide crystal. The GRD performances for different crystal types, data acquisition modes, working modes, and incident positions are also analyzed in detail. We show that the calibration campaign is comprehensive, and the calibration results are generally consistent with simulations as expected.

**GRD detector, ground-based calibration, energy response, detection efficiency**

## 1 Introduction

Gravitational wave high-energy Electromagnetic Counterpart All-sky Monitor project (GECAM) is a small space scientific exploration project which focuses on detecting high-energy electromagnetic counterparts of gravitational waves and other gamma-ray transients such as gamma-ray bursts and soft gamma-ray repeaters.

GECAM is composed of two identical microsatellites. The payload of each satellite includes 25 gamma-ray detectors (GRDs), 8 charged particle detectors (CPDs) and a payload electronics processor (EBOX). The GECAM satellite diagram is shown in Figure 1.1.

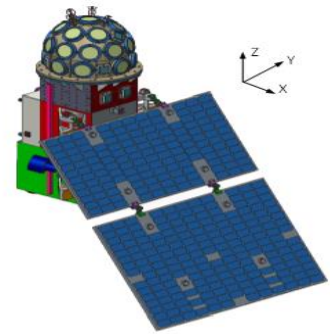

**Figure 1.1** Illustration of a GECAM satellite. The payload coordinate is shown.

These 25 GRDs and 8 CPDs point to different directions and are nearly evenly distributed in the solid angle (except for the direction facing the earth). The GRD detector is the main detector of the GECAM satellite payload. GRD is a high detection efficiency detector which has a large area and large field of view. Its main function is to trigger, locate, measure the light curve and spectrum of gamma-ray bursts from about 6 keV to 5 MeV. The schematic diagram of the layout of the GECAM satellite detector is shown in Figure 1.2.

*Corresponding author (email: hejianjian@ihep.ac.cn, anzh@ihep.ac.cn, pengwx@ihep.ac.cn, lixq@ihep.ac.cn, xiongsl@ihep.ac.cn)

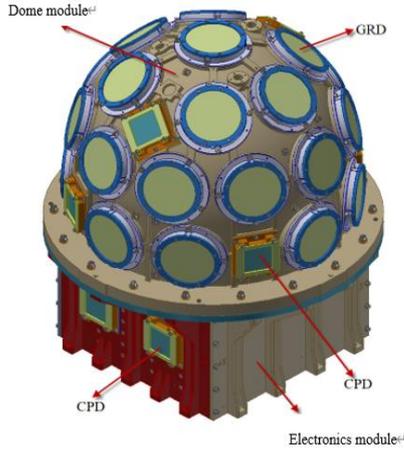

**Figure 1.2** Schematic diagram of the layout of the GECAM satellite detector.

To meet the scientific needs of GECAM and the requirements of the small satellite platform, the GRD employed the novel design of 3-inch LaBr3 crystal readout by a dedicated SiPM array. Lanthanum bromide (LaBr3) scintillation crystal has the advantages of high light yield, high density and fast decay time. The silicon photomultiplier tube (SiPM) features a simple and compact structure and easy miniaturization. The design of the GRD detector is shown in Figure 1.3. From top to bottom, there are crystal box components, silicone pad, reflective film, SiPM board, mechanical structure. Gamma-rays are incident from the beryllium window of the crystal box assembly and detected by the lanthanum bromide crystal. The scintillation light photons induced by the energy deposition of incident gamma-rays pass through the quartz window and are received by the SiPM. SiPM converts the very weak light signal to the measurable electrical signal, which is further amplified by the preamplifier. The amplified signal is finally connected to the data acquisition circuit board where the signal is digitized and transferred to the data management board of GECAM payload.

## 1.1 Ground-based calibration of GRD

The calibration process of the detector is to establish the relationship between the detected output signal and the input signal (gamma-rays) to ensure that the input can be accurately derived from the output measurement when the satellite operates on the orbit. The ground-based calibration items of GRD detectors can be divided into energy response, detection efficiency, spatial response, temperature bias response, time characteristic and space environment response. Among them, the energy response and detection efficiency calibration are the most important parts of GRD calibration. To obtain the energy response and detection efficiency curve, it is necessary to select multiple energy points of X-rays as the calibration source. Calibration sources used for ground-based calibration include X-ray beam-based single and dual crystal monochromators and radioactive sources.

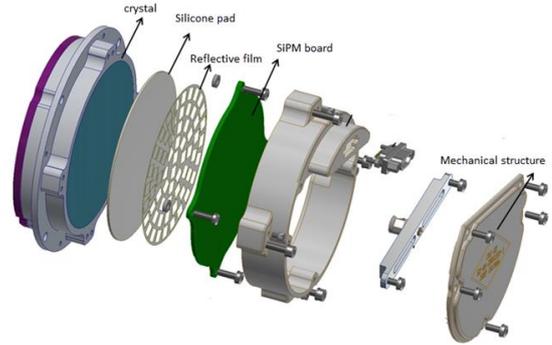

**Figure 1.3** Exploded view of the structure of the gamma-ray detector (GRD).

Ground-based calibration requires that the measurement points should cover 8 keV-2 MeV as much as possible, the total measurement points should be more than 20, the statistics of each energy point should be no less than 100,000, and individual detectors are needed to measure the absorption edge in a fine energy step ($\leqslant$0.1 keV/step). This calibration low-energy band (8-160 keV) is calibrated with the X-ray beam of the National Institute of Metrology China, and the high-energy band (greater than 160 keV) is calibrated with radioactive sources.

Ground-based calibration of GRD low-energy band is mainly carried out on the calibration device of the National Institute of Metrology China. During the calibration process, the diameter of the primary collimating aperture is 9 mm, the diameter of the secondary collimating tube's entrance aperture is 6 mm, and the exit aperture is 3 mm. During the calibration process, the environment ambient temperature is controlled at 20～25℃, and the temperature fluctuation is < ±2℃. The focus is on the energy response and detection efficiency of the GRD under 8～160 keV X-ray beams. Because lanthanum bromide crystals have absorption edges at 13-15 keV and 38-41 keV (Ivan V Khodyuk & Pieter Dorenbos 2010), and lanthanum bromide crystals have obvious energy nonlinearities less than 15 keV (Ivan V Khodyuk & Pieter Dorenbos 2010; Zhang 2019), it is necessary to perform detailed calibration at these energy points to study the nonlinearity of the response of GRD, and to perform refined measurements around the absorption edge of the lanthanum bromide crystal. With the increase of X-ray energy, the performance of the GRD changes slowly, and the number of energy points for calibration can be appropri-ately reduced to save calibration time. Because it is tested at room temperature, the noise of the detector (mostly for SiPM) is larger than the designed working temperature (about -20℃), and the test energy point starts from 7.5-10 keV. Table 1.1 shows the main design performance and calibration test plans.

**Table 1.1**　Main design technical indicators and calibration test plan of GRD.

| Items | Performance | Calibration test plan |
|---|---|---|
| Energy | 8 keV∼2 MeV | X-ray beam combined with radioactive source test |
| Area | > 40 cm² per det | given by crystal size |
| Dead time | ≤5 μs | given by other calibration items |
| Energy resolution | <18%@59.5 keV | given by the radioactive source of Am-241 |
| Efficiency | >50%@8 keV | X-ray beam combined with Fe-55 efficiency test |

## 1.2　Ground-based calibration facility for GRD

The ground-based calibration test of the payload GRD detector was carried out in the National Institute of Metrology China. Its X-ray ground-based calibration device HXCF was specifically established for the calibration of the high-energy detector of the hard X-ray modulation telescope (HXMT) (Zhou 2014). The structure of HXCF includes six components: X-ray machine, dual-crystal monochromator (or single crystal monochromator), collimation system, support system, shielding device, and control equipment. The X-ray beam generates continuous X-rays. The dual-crystal monochromator (or single crystal monochromator) makes the X-ray monochromatic through Bragg diffraction; there are two collimators with lead shielding in the X-ray path, The collimator port is equipped with a beam-limiting diaphragm to limit the spot size; the position of the detector can be adjusted by a translation stage and a rotation stage. The schematic diagram of the calibration device is shown in Figure 1.4 and Figure 1.5 (Song 2019; Zhou 2019).

For this study, the dual crystal calibration device uses two Si220 crystals with an energy range of 35-75 keV and two Si551 crystals with an energy range of 80-160 keV. In addition, there is a set of single crystal calibration device, using a LiF200 crystal, the energy range is 7-40 keV. Due to the modular design of GECAM satellites, each satellite contains 25 GRD detectors, with a total of 50 installed GRDs for two satellites. During the ground-based calibration test, 67 GRDs were subjected to more than 80 single-doped crystal tests and more than 140 double-doped crystal tests.

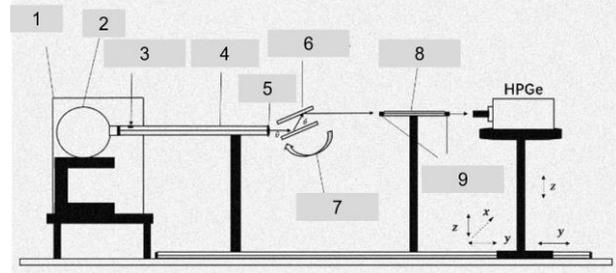

**Figure 1.4**　Schematic diagram of the calibration device of the dual-crystal monochromator based on the X-ray beam（1. Shield box, 2. Light machine, 3. Laser collimator, 4. Front collimator, 5. Light, 6. Crystal monochromator, 7. Rotary table, 8. Rear collimation tube, 9. Light）.

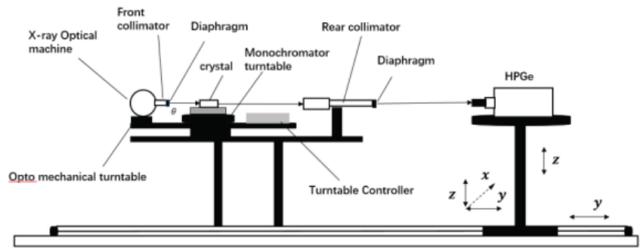

**Figure 1.5**　Schematic diagram of single crystal monochromator calibration device based on X-ray machine.

**Table 1.2**　Performance parameters of X-ray machine monochromator calibration device.

| Items | Technical index |
|---|---|
| Energy | LiF200：7~40 keV<br>Si220：35~75 keV<br>Si551：80~160 keV |
| Monochromatic light ratio | ＞90% |
| Monochromatic luminous flux | ＞200cts/s |
| Spot size | Adjustable diameter（3 mm）, customizable |
| Energy divergence (σ) | <1.2%@59.5 keV |
| Beam line position | deviation＜2.4 mm@φ2 mm |

## 2　Calibration of the HPGe detector

For the measurement of performance parameters of monochromatic X-ray sources, including energy, monochromaticity, photon intensity, etc. There are a variety of X-ray detectors to choose from, high-purity germanium detector (HPGe) is used in the calibration process of GRD detectors. The high-purity germanium detector is used as a standard detector for measuring the energy, monochromaticity and absolute flux of the monochromatic X-ray sources.

Corresponding to the calibration of each energy point of each GRD, it is necessary to use the HPGe detector to test the beam before the test to obtain the accurate energy value from the X-ray beam through the crystal monochromator.

Before the entire calibration work is carried out, it is necessary to calibrate the energy of the HPGe detector. The types of radioactive sources used for calibration include Fe-55, Am-241, Co-57 and Ba-133. As shown in Table 2.1, 5.90 keV characteristic peak is chosen at the spectrum of Fe-55, 26.34 keV and 59.54 keV two characteristic peaks are chosen at Am-241's spectrum, Co-57 selects 4.40 keV, 14.41 keV, 122.06 keV and 136.47 keV. Ba-133 selects 81.00 keV characteristic peak, a total of eight energy points is used for energy calibration of HPGe detector.

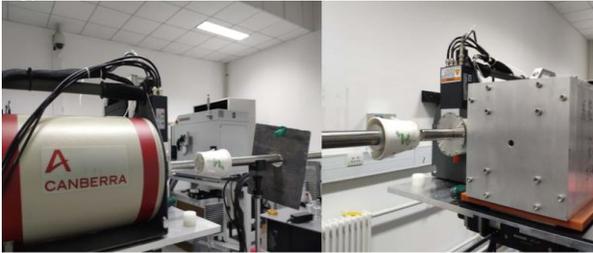

**Figure 2.1**　Use of radioactive sources to calibrate and test high-purity germanium detector.

On June 24, 2020, the energy calibration of the high-purity germanium detector was carried out with the above four radioactive sources. In the calibration process, two high-purity germanium detectors are involved, called the old and the new respectively. The calibration of dual crystals uses new high-purity germanium detector. The calibration scene of the new and old high-purity germanium detectors is shown in Figure 2.2.

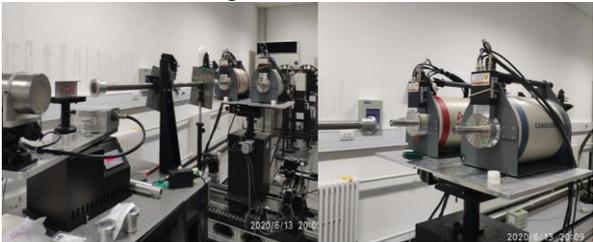

**Figure 2.2**　Comparison calibration site of new and old high-purity germanium (red: old high-purity germanium detector, blue: new high-purity germanium detector).

The spectrum and characteristic peaks of the above four radioactive sources are shown in Figure 2.3 and 2.4. Perform Gaussian fitting of each characteristic peak to obtain the corresponding peak position and broadening σ (as shown in Table 2.1). According to the peak position corresponding to each energy point, linear function fitting is performed to obtain the E-C calibration relationship of the two high-purity germanium detectors, as shown in Figure 2.5. In the process of GRD ground-based calibration using the X-ray beam, according to the E-C calibration relationship of the high-purity germanium detector, the accurate energy value of the beam after the monochromator can be obtained, which provides energy information for the subsequent energy response calibration of the GRD detector. In addition, according to the spectral broadening σ measured by the high-purity germanium detector, the energy resolution of the high-purity germanium detector can be obtained, and the detection efficiency curve of the GRD can be obtained according to the peak area of the universal peak. The energy resolution obtained by using the radioactive source in this chapter can be used as the intrinsic resolution of the high-purity germanium detector, denoted as $Energy\ Resolution\_radi = \frac{2.355\sigma}{Peak}$, as shown in Figure 2.6.

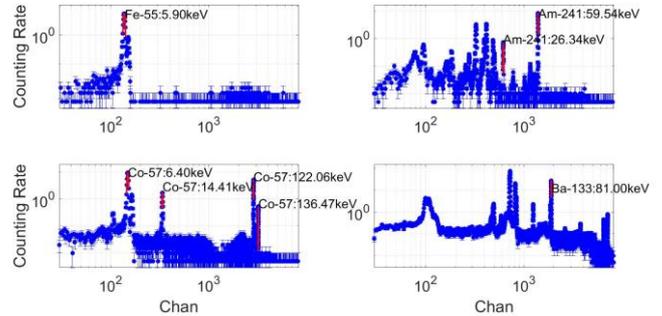

**Figure 2.3**　Energy spectrum and characteristic peaks of radioactive sources such as Fe-55, Am-241, Co-57 and Ba-133: old high-purity germanium detector.

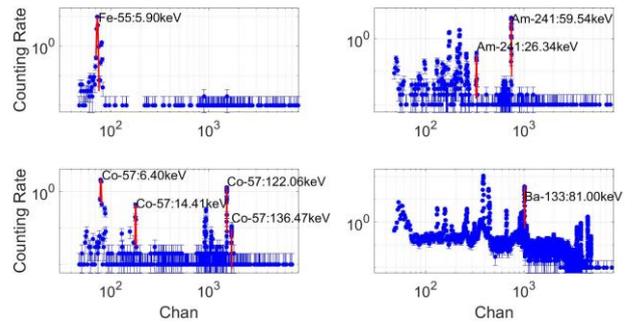

**Figure 2.4**　Energy spectrum and characteristic peaks of radioactive sources such as Fe-55, Am-241, Co-57 and Ba-133: new high-purity germanium detector.

Figure 2.5 shows E-C relationship of the new and old high-purity germanium detectors, as well as fitting with a linear function to obtain the relationship, which is mainly used for the subsequent use of the X-ray beam to perform the energy response of the GRD detector calibration. Figure 2.6 (left) shows the energy resolution of the new and old high-purity germanium detectors. It can be seen from the figure that the energy resolution of the new high-purity germanium detector is better than that of the old high-purity germanium detector. This phenomenon is more obvious at the low-energy end. As the energy increases, the energy resolution of the two high-purity germaniums tends to be the same. As the intrinsic energy resolution of the high-purity germanium detectors, these will be mainly used in the calculation of the energy resolution of the subsequent GRD detector.

**Table 2.1** Types of radioactive sources involved in the energy response calibration of high-purity germanium detectors and their characteristic peak Gaussian fitting results.

| Old HPGe-Det | | | | | |
|---|---|---|---|---|---|
| Radioactive source | Energy（keV） | Peak | err | σ | err |
| Fe-55 | 5.90 | 137.32 | 0.03 | 2.38 | 0.03 |
| Am-241 | 26.34 | 614.30 | 0.09 | 2.95 | 0.08 |
| | 59.54 | 1388.90 | 0.03 | 3.77 | 0.03 |
| Co-57 | 6.40 | 149.29 | 0.05 | 2.23 | 0.05 |
| | 14.41 | 336.06 | 0.09 | 2.49 | 0.10 |
| | 122.06 | 2848.80 | 0.07 | 4.98 | 0.07 |
| | 136.47 | 3186.00 | 0.15 | 5.17 | 0.11 |
| Ba-133 | 81.00 | 1889.30 | 0.03 | 4.38 | 0.03 |
| New HPGe-Det | | | | | |
| Radioactive source | Energy（keV） | Peak | err | σ | err |
| Fe-55 | 5.90 | 73.12 | 0.06 | 0.85 | 0.04 |
| Am-241 | 26.34 | 328.54 | 0.19 | 1.42 | 0.15 |
| | 59.54 | 742.80 | 0.08 | 1.85 | 0.06 |
| Co-57 | 6.40 | 79.42 | 0.02 | 0.87 | 0.02 |
| | 14.41 | 179.44 | 0.01 | 1.09 | 0.01 |
| | 122.06 | 1523.70 | 0.11 | 2.60 | 0.09 |
| | 136.47 | 1703.80 | 0.37 | 2.75 | 0.33 |
| Ba-133 | 81.00 | 1010.70 | 0.01 | 2.15 | 0.01 |

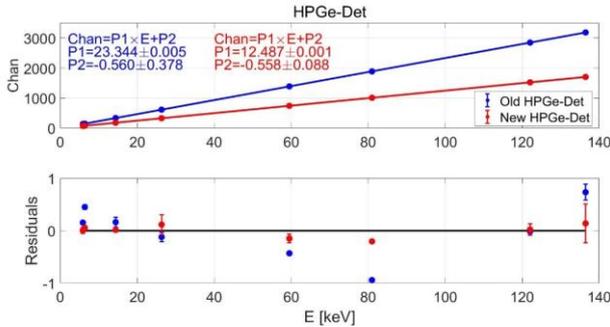

**Figure 2.5** E-C relationship of high-purity germanium detector: the blue data point is the old high-purity germanium data, and the red data point is the new high-purity germanium data (the ordinate Chan represents the track address recorded by the data acquisition board, and Residual is defined as Residual= measured value-model value, the same below).

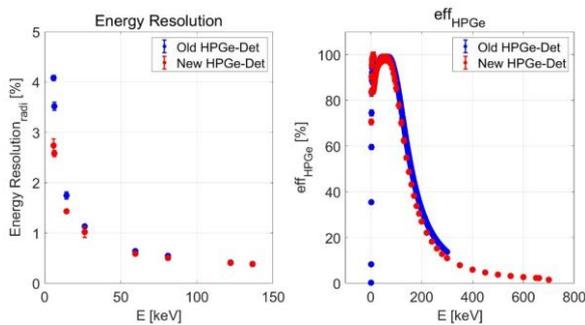

**Figure 2.6** Intrinsic energy resolution of high-purity germanium detectors (left), detection efficiency (right): the blue data points are the old high-purity germanium data, and the red data points are the new high-purity germanium data (the ordinate on the left is Intrinsic energy resolution, the ordinate on the right is the detection efficiency)

In addition, the detection efficiency of the high-purity germanium detectors are given by the measured values and the simulated values. The detection energy range of the HPGe detector is within 350 keV. During the calibration process, select different characteristic peaks of radioactive sources suitable for the detection range of the HPGe detector to calibrate the detectors (respectively Am-241(59.54 keV), Cd-109(21.99 keV/22.16 keV/88.03 keV), Co-57(14.41 keV/122.06 keV/136.47 keV)). And then efficiency values of other energy points are given by Monte Carlo simulation (Liu 2016). The full energy peak detection efficiency curve of the new and old high-purity germanium detectors are shown in Figure 2.6 (right). The detection efficiency of the old high-purity germanium detector is slightly better than that of the new high-purity germanium above 100 keV.

# 3 Energy response and detection efficiency calibration

According to the technical process for the development of the GECAM satellite prototype, the GRD detector will be calibrated on the ground after the assembly is completed and the performance test and quality assurance are carried out. The pictures of GRD are shown in Figure 3.1.

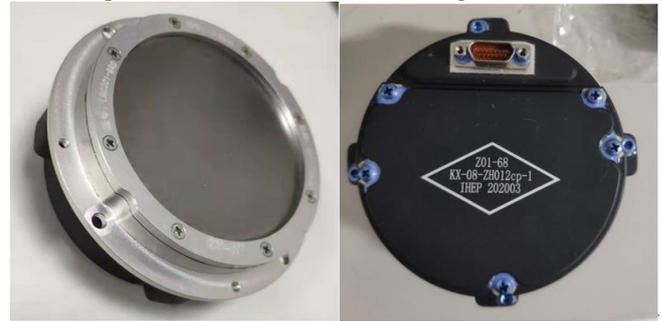

**Figure 3.1** The appearance of official version of the GRD detector.

For each GRD, the energy response and detection efficiency can be calibrated through statistics, comparison, and integration of the spectrum change, counting rate, time interval spectrum, temperature curve, baseline distribution, etc. of each test.

## 3.1 E-C relationship

In this chapter, we select 5 GRDs (respectively Z01-10, Z01-70, Z01-40, Z01-44, Z01-53) single crystal and dual crystal test data for analysis (including scientific data and telemetry data). Among them, the high gain channel (marked with HighGain) contains single crystal and dual crystal test data, and the energy covers 8-160 keV, and the low gain channel (marked with LowGain) is dual crystal test data, and the energy range is 40-160 keV. Fit the energy spectrum measured under the beam line from X-ray beam of the GRD, and fit its full energy peak with a Gaussian function. Figure 3.2 shows the fitting results of the full energy peak of the five GRDs (high gain channel 28 keV, low gain channel 80 keV/79 keV). Through the Gaussian fitting of the a full energy peak, the peak value, broadening σ, and peak area of the full energy peak can be obtained $Area_{GRD}$, and the E-C relationship scatters plot can be obtained by statistical analysis of the peak position corresponding to each energy points (as shown in Figure 3.3). It can be seen from Figure 3.3 that the GRD detection energy has a good linear relationship with the recorded Chan of the data acquisition board, and there are differences in the linear relationship between the GRDs. This difference involves the consistency of the GRDs and will be discussed in a subsequent articles.

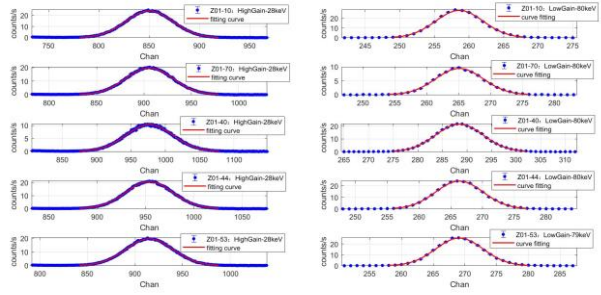

**Figure 3.2** Fitting of full energy peaks in the energy spectrum of GRDs: the ordinate is the counting rate, and the abscissa is Chan recorded by the data acquisition board (left: high-gain channel, right: low-gain channel).

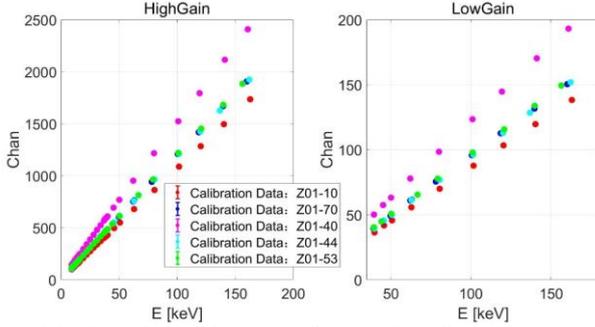

**Figure 3.3** E-C relationship scatters diagram: the ordinate Chan recorded by the data acquisition board, the abscissa E is the energy (left: high gain channel, right: low gain channel).

Select the E-C scatter data of the Z01-40 for a quadratic function fitting, and perform segmentation (13.48±1 keV, 38.89±1 keV) near the two absorption edges of the lanthanum bromide crystal to obtain the E-C relationship of high/low gain channel. The E-C relationship fitting results are shown in Figure 3.4-3.5, which shows that the quadratic function model can basically meet the energy calibration requirements. The E-C relationship of the high gain channel is

$Chan = 0.739 \times E^2 + 1.026 \times E + 71.509$     $(E < 12.48 \; keV)$,
$Chan = -0.018 \times E^2 + 16.283 \times E - 9.386$
     $(14.48 \; keV < E < 37.89 \; keV)$,
$Chan = -0.002 \times E^2 + 15.252 \times E + 6.994$
     $(E > 39.89 \; keV)$.

The E-C relationship of the low gain channel is

$Chan = 1.200 \times E + 3.45$     $(E > 39.89 \; keV)$.

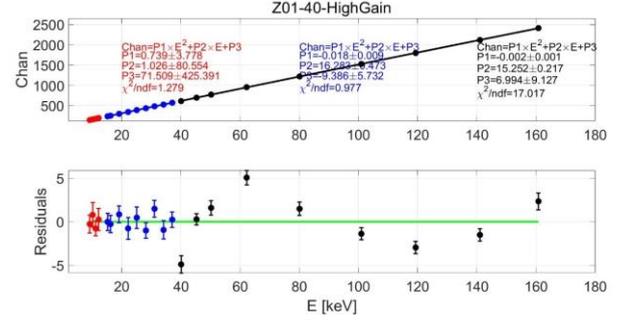

**Figure 3.4** E-C relationship fitting result of high gain channel: Z01-40.

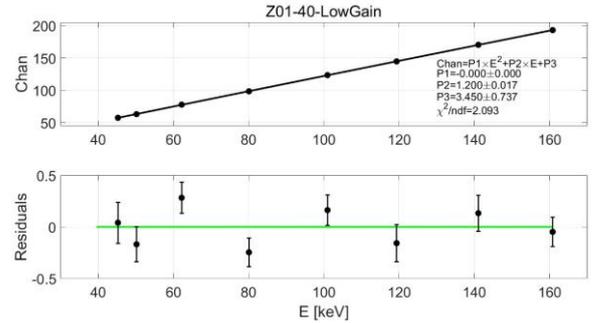

**Figure 3.5** E-C relationship fitting result of low gain channel: Z01-40.

## 3.2 Energy resolution

According to the energy spectrum analysis mentioned above, the full energy peak position and spectrum broadening σ can be obtained, and then the energy resolution corresponding to each energy point of the GRD can be calculated (denoted as $Energy \; Resolution_{GRD}$). When using an X-ray beam to calibrate the energy resolution of the GRD, the result obtained by this method is not the intrinsic energy resolution (denoted as $Energy \; Resolution$), it covers the effect of beam (denoted as $Energy \; Resolution_{beam}$) and the effect of the intrinsic energy resolution. The calculation formula for the intrinsic energy resolution of the GRD is as follows

$$Energy \; Resolution = \frac{2.355\sigma}{Peak_0}. \qquad (1)$$

Where $\sigma = \sqrt{\sigma_{GRD}^2 - \sigma_{beam}^2}$, $\sigma_{GRD}$ is the broadening of the full energy peak measured by GRD, which can be obtained by GRD energy spectrum fitting. $\sigma_{beam} = \sqrt{\sigma_{HPGe}^2 - \sigma_{radi}^2}$ is the beam line broadening of the X-ray beam, $\sigma_{HPGe}$ is the X-Ray beam full peak broadening measured by the high-purity germanium detector, and $\sigma_{radi}$ is the radioactive source full peak measured by the high-purity germanium detector. $\sigma_{HPGe}$ can be obtained by fitting the HPGe energy spectrum, and $\sigma_{radi}$ is used to obtain the numerical values at various energies through model fitting based on several characteristic peaks broadening measured by high-purity germanium in Chapter 2. $Peak_0$ is the peak position value after deducting the baseline. Figure 3.6 shows the broadening of the full energy peak measured by the high-purity germanium detector and the broadening of the full energy peak measured by the GRD detector at the same energy point of 28 keV.

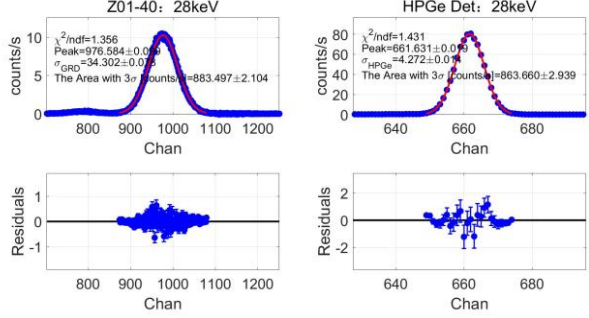

**Figure 3.6** Full energy peak broadening measured by GRD ($\sigma_{GRD}$) vs Full energy peak broadening measured by high-purity germanium ($\sigma_{HPGe}$) at the point of 28 keV: X-ray beam.

As described in section 3.1, we select the single crystal and double crystal test data of the same 5 GRDs (respectively Z01-10, Z01-70, Z01-40, Z01-44, Z01-53), and perform energy spectrum analysis on them respectively. We obtain peak positions and broadening, and then calculate the intrinsic energy resolution corresponding to each energy point according to formula 1. The scatter plot of the energy resolution of this batch of samples as a function of energy is shown in Figure 3.7.

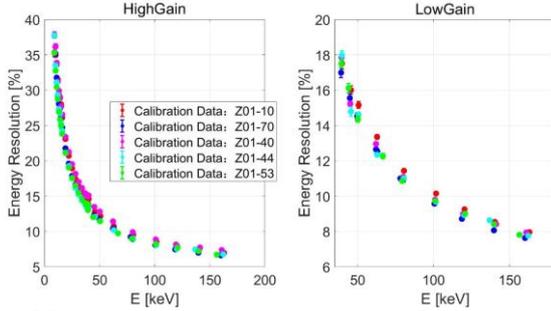

**Figure 3.7** E-ER relationship scatter diagram (left: high gain channel, right: low gain channel, abscissa is energy, ordinate is energy resolution).

Select the E-ER data of the Z01-40 for model fitting, and similarly perform segmentation (13.48±1 keV, 38.89±1 keV) near the two absorption edges of the lanthanum bromide crystal, the fitting result of the E-ER relationship, as shown in Figure 3.8-3.9

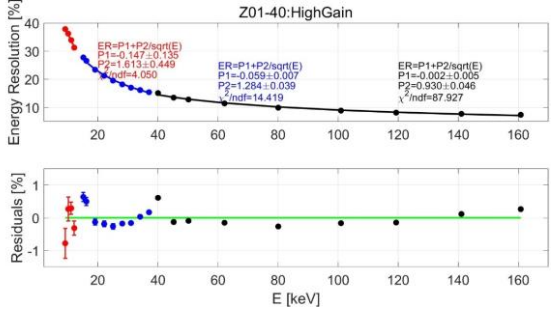

**Figure 3.8** E-ER relationship fitting result of high gain channel: Z01-40.

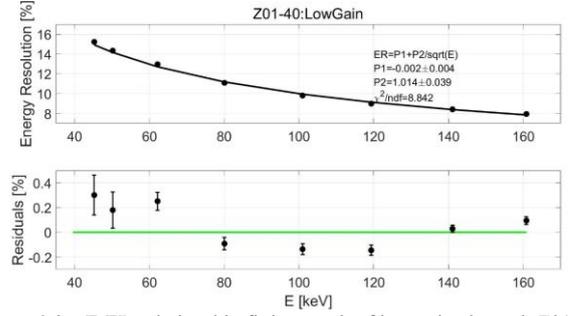

**Figure 3.9** E-ER relationship fitting result of low gain channel: Z01-40.

### 3.3 Detection efficiency

The detection efficiency of GRD can be indirectly given by the efficiency of the high-purity germanium detector. Suppose the monochromatic light intensity $I(E)$ is expressed as $I(E) = \frac{Area_{HPGe}(E)}{eff_{HPGe}(E)}$, where $Area_{HPGe}(E)$ is the area (counting rate) of the full energy peak at the point of E, measured by the high-purity germanium detector, $eff_{HPGe}(E)$ is the detection efficiency of the full energy peak of the HPGe detector for X/γ rays with energy E, combined by the Monte Carlo simulation and the measured data is given in Figure 2.6 (right).

The detection efficiency of the GRD detector for the full energy peak at the point of E can be expressed as $eff_{GRD}(E) = \frac{Area_{GRD}(E)}{I(E) \cdot K_I}$, where $K_I$ is the stability factor of beam from X-ray machine, $Area_{GRD}(E)$ is the area (counting rate) of full energy peak measured by GRD detector at energy E. From the above two formulas, the expression of the detection efficiency of the GRD detector can be derived as formula 2,

$$eff_{GRD}(E) = \frac{eff_{HPGe}(E)}{Area_{HPGe}(E) \cdot K_I} Area_{GRD}(E). \qquad (2)$$

The error of GRD detection efficiency mainly comes from the following parts: (1) The statistical error derived from the energy spectrum measured by the high-purity germanium detector and the fitting error of the full energy peak: fitting the full energy peak of the response spectrum can give the fitting error of the peak position and the broadening σ, which can be used to obtain the peak area $Area_{HPGe}(E)$. $Area_{HPGe}(E) = \sum_i \frac{N_i}{t}$, where $N_i$ is the count number of the i-th channel ($i = peak \pm 3\sigma$), t is the live time, and the error of peak area is $\sigma_{Area_{HPGe}(E)} = \sqrt{\frac{\sum_i(\sigma_{N_i})^2}{t}}$, where $\sigma_{N_i} = \frac{\sqrt{N_i}}{t}$. The uncertainty of the detection efficiency of the full energy peak of the HPGe detector in the entire energy range is $\frac{\sigma_{eff_{HPGe}(E)}}{eff_{HPGe}(E)} = 1.8\%$ (Liu 2016). (2) The statistical error derived from the energy spectrum measured by the GRD detector and the fitting error of the full energy peak: fitting the full energy peak of the response spectrum can give the fit-ting error of the parameter peak position and the spreading σ, and similarly, use it to obtain the peak area $Area_{GRD}(E)$. $Area_{GRD}(E) = \sum_i \frac{N_i}{t}$, where $N_i$ is the count number of the i-th channel ($i = peak \pm 3\sigma$), t is the live time, and the error of peak area error is $\sigma_{Area_{GRD}}(E) = \sqrt{\sum_i (\sigma_{\frac{N_i}{t}})^2}$, where $\sigma_{\frac{N_i}{t}} = \frac{\sqrt{N_i}}{t}$.

(3) The error introduced by the instability of the X-ray beam is quantified by the $K_I$ factor, $K_I = \frac{CR_{GRD}}{CR_{HPGe}}$, where $CR_{GRD}$ is the beam counting rate when the GRD is calibrated, and $CR_{HPGe}$ is the beam counting rate when the high-purity germanium detector is used for measurement, $CR_{GRD}$ and $CR_{HPGe}$ can be obtained by model deduction based on the light curve measured by GRD according to the time difference between two experiments.

Same as Section 3.1 and 3.2, select single crystal and double crystal test data of 5 GRDs (respectively Z01-10, Z01-70, Z01-40, Z01-53) for analysis, the detection efficiency varies with energy is shown in Figure 3.10-3.11.

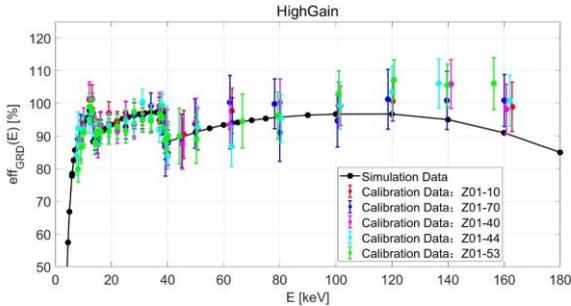

**Figure 3.10** Measured detection efficiency of high gain channel vs. simulation efficiency: the black dotted line in the figure is the simulated efficiency curve of the GRD, and the colored scatter points are the measured efficiency of the GRDs.

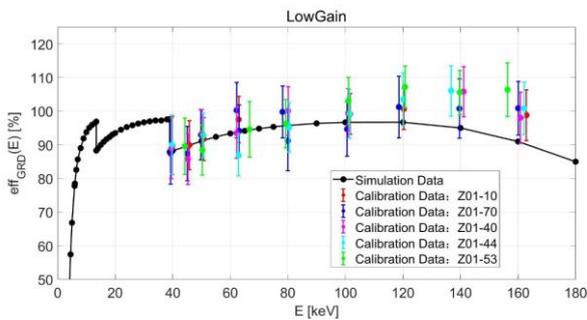

**Figure 3.11** Measured detection efficiency of low gain channel vs. simulation efficiency: the black dotted line in the figure is the simulated efficiency curve of the GRD, and the colored scatter points are the measured efficiency of the GRDs.

It can be seen from Figure 3.10-3.11 that the measured efficiency of the GRDs are basically consistent with the detection efficiency obtained by MC simulation, only the measured value is higher than the simulated value at about 160 keV. According to the second chapter, the detection efficiency of high-purity germanium is determined by the measured values and the values from MC simulation, the measured value only involves 7 energy points of 14.41 keV,

21.99 keV, 22.16 keV, 59.54 keV, 88.03 keV, 122.06 keV, 136.47 keV, etc., lacking the actual measurement near 160 keV, so the efficiency at 160keV is higher than the simulated value is probably caused by the inaccurate simulation values of the high-purity germanium detector at this energy point.

# 4 Energy response and detection efficiency around the absorption edges of the lanthanum bromide crystal

It is known that there are two absorption edges in lanthanum bromide crystals (Ivan V Khodyuk & Pieter Dorenbos 2010), namely the absorption edge of bromine (13.48 keV) and the absorption edge of lanthanum (38.89 keV). Select the energy points near the two absorption edges and take 0.1 keV as a step, then we can get the finer measurement results around the absorption edges.

## 4.1 E-C relationship

Choose the Z01-08 as the sample, select its single crystal experiment and double crystal experiment data for analysis. The wide energy range and E-C scatter diagrams near the absorption edge are shown below. In order to make the situation near the absorption edges clearer, make E-Chan/E relationship diagrams in Figure 4.1. Fit the measurement points near the absorption edges to a quadratic function, and obtain the E-C relationship near the absorption edges as shown in Figure 4.2-4.3.

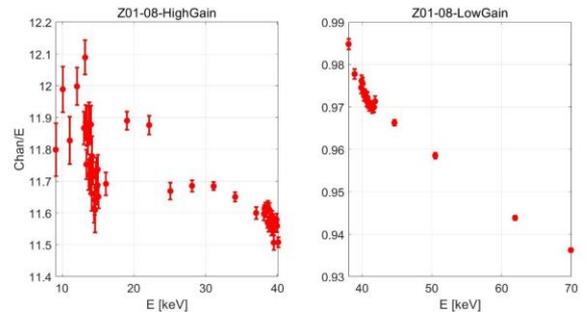

**Figure 4.1** Precision measurement results near the absorption edges of the Z01-08: E-Chan/E relationship scatter diagram, the left picture is the data of the high gain channel, and the right picture is the data of the low gain channel.

According to the comparison between Figure 3.4-3.5 and Figure 4.2-4.3, it can be seen that the E-C relationship near the absorption edge is significantly different from the E-C relationship of other energy segments.

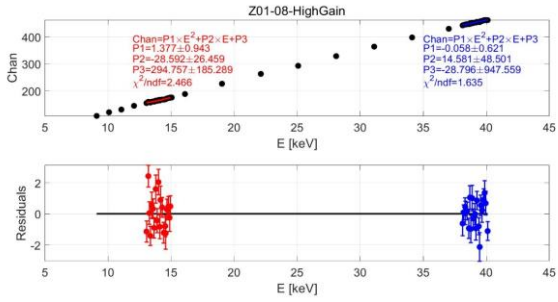

**Figure 4.2**  Absorption edges precision measurement results: E-C relationship fitting results of Z01-08 high-gain channel.

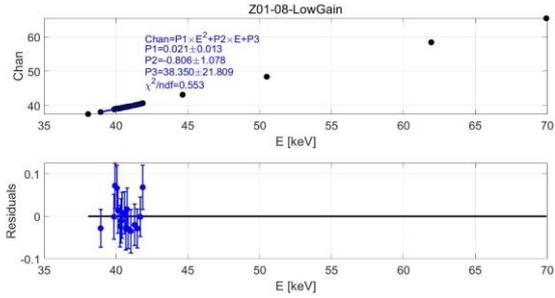

**Figure 4.3**  Absorption edges precision measurement results: E-C relationship fitting results of Z01-08 low-gain channel.

## 4.2  Energy resolution

According to the energy resolution calculation method described in section 3.2, the relationship between the energy resolution near the absorption edge and the energy E can be obtained (Figure 4.4). Fit the measurement points near the absorption edges to the model, and obtain the E-ER relationship curve near the absorption edges as shown in Figure 4.5-4.6.

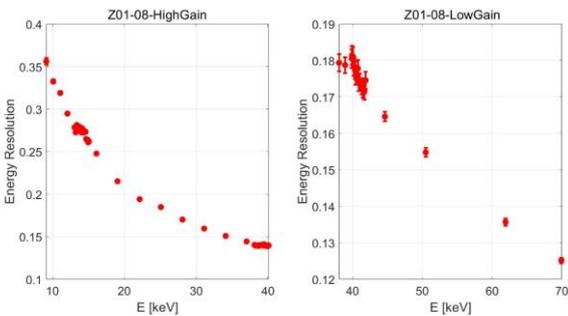

**Figure 4.4**  The precision measurement results near the absorption edges of the Z01-08: E-ER relationship (left: high-gain channel, right: low-gain channel).

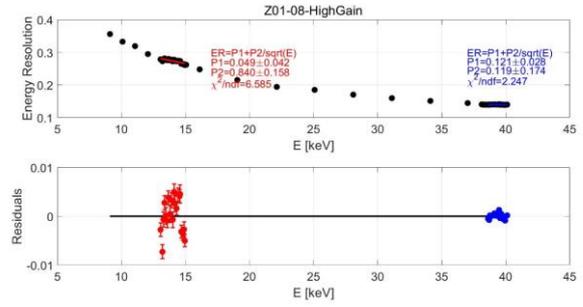

**Figure 4.5**  Absorption edges precision measurement results of Z01-08 high-gain channel: E-ER relationship fitting results.

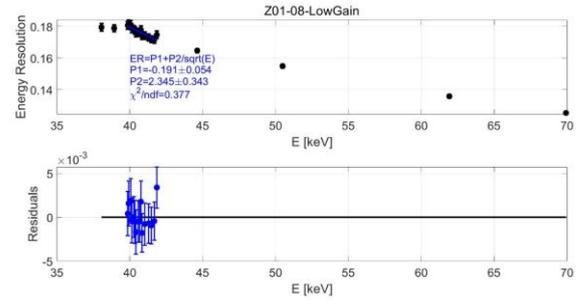

**Figure 4.6**  Absorption edges precision measurement results of Z01-08 low-gain channel: E-ER relationship fitting results.

According to Figure 3.8-3.9 and Figure 4.5-4.6, it can be seen that the E-ER relationship near the absorption edges is significantly different from the E-ER relationship of other energy segments.

## 4.3  Detection efficiency

According to the calculation method of detection efficiency described in section 3.3, the detection efficiency corresponding to each energy point near the absorption edge can be obtained. Figure 4.7-4.8 show the detection efficiency of the high gain channel and the detection efficiency of the low gain channel of Z01-08 respectively, and compared with the Monte Carlo simulation values. The measured values and the simulation values are consistent within the error range, indicating that the physical model of GRD detector is credible.

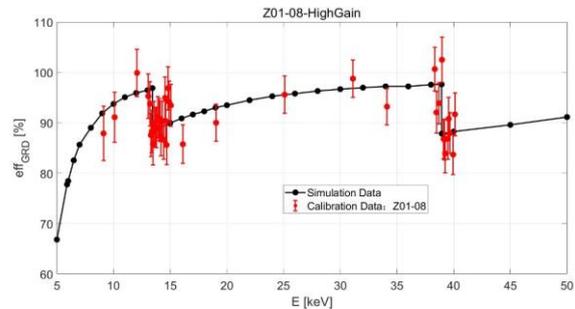

**Figure 4.7**  High-gain channel absorption edge measurement results of Z01-08: detection efficiency.

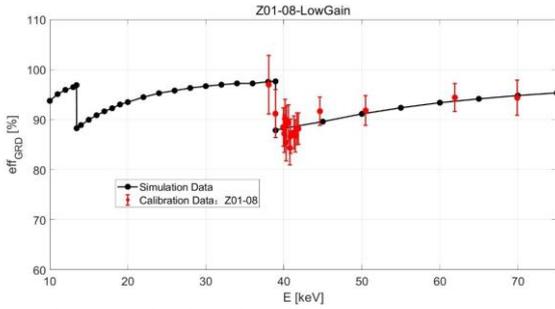

**Figure 4.8** Low-gain channel absorption edge measurement results of Z01-08: detection efficiency.

# 5 Performances studies

## 5.1 Crystal doping type

Since the GRD detectors are produced in batches by Beijing Glass Research Institute Co., Ltd, the lanthanum bromide crystals of each batch are slightly different. According to the crystal differences of each batch, the classification standard is developed to divide the lanthanum bromide crystals into single-doped, double-doping and middle-doping, which are mainly caused by the different proportions of cerium (Ce) ions and strontium (Sr) ions doping. Among the 50 GRDs installed, there are 37 single-doped lanthanum bromide crystals, 11 double-doped lanthanum bromide crystals, and 2 middle-doped intermediate states. According to the existing research results, the nonlinear response of the low-energy end energy of different doped crystals is slightly different (I. V. Khodyuk 2013). Under different doping types, lanthanum bromide crystals exhibit significantly different energy nonlinear responses, and this nonlinear characteristic is widely present in the crystals used by multiple detectors, such as Sodium iodide crystals used in the HXMT satellite show similar nonlinear characteristics (Li 2020).

Differentiate different crystal types, namely single-doped crystals, middle-doped crystals, and double-doped crystals. The data of 50 installed GRDs are statistically analyzed and normalized at 160 keV. Since different crystal types have obvious nonlinear phenomena mainly in the low-energy region, here we only select the data of the high gain channel as samples for analysis, and get the scatter diagram of the Chan/E-E relationship (Figure 5.1). It can be seen that different crystals Below 15 keV, the E-C relationship shows a very obvious difference, and this difference can even be extended to 40 keV. In order to study the effect of crystal type on energy resolution and detection efficiency, Z01-29 is selected as the sample for single-doped crystal, Z01-35 is selected for middle-doped crystal, and Z01-70 is selected as sample for double-doped crystal. The energy resolution and detection efficiency are shown in Figure 5.2-5.3. According to Figure 5.2, the energy resolution of single-doped crystals is overall inferior to that of middle-doped intermediate crystals, and the energy resolution of middle-doped crystals is overall worse than that of double-doped crystals, this difference tends to decrease with increasing energy. In terms of detection efficiency, the measured detection efficiency values of the three doped crystals all match the simulated values well.

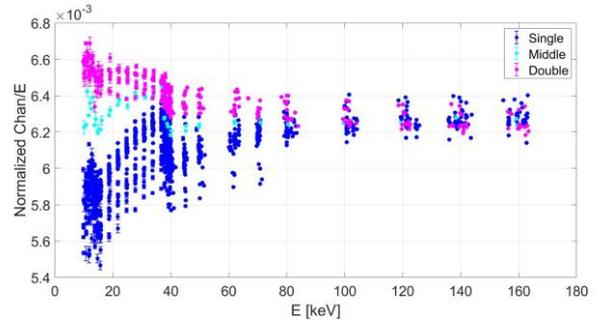

**Figure 5.1** E-C relationship scatter diagram of different crystal types (high gain channel): Single means single-doped crystal, Middle means middle-doped crystals, Double means double-doped crystals.

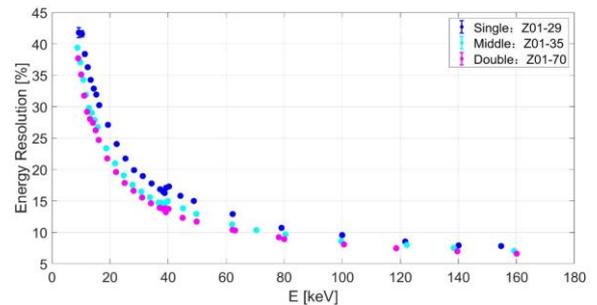

**Figure 5.2** Energy resolution scatter diagram of different crystal types (high gain channel): single- doped crystal Z01-29, middle-doped crystal Z01-35, double-doped crystal Z01-70.

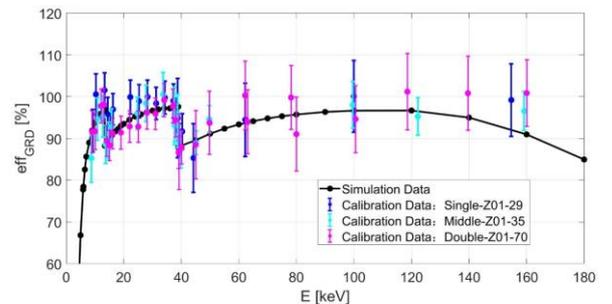

**Figure 5.3** Scatter diagram of detection efficiency of different crystal types (high gain channel): single-doped crystal Z01-29, middle-doped crystal Z01-35, double-doped crystal Z01-70.

The energy nonlinear response phenomenon of different crystals has been mentioned many times in the literature. In Alekhin's 2013 article, I. V. Khodyuk et al. (2013) discussed how to improve the energy resolution of lanthanum bromide crystals for detecting gamma rays, and gave the X-ray response curve of different crystals: The energy response of different crystals is different, even for the same crystal LaBr3, the difference of doping type will cause obvious differences in the response curve.

The energy nonlinear response of the crystal described above will have a significant impact on the calibration of the energy response of the detectors: on the one hand, the energy nonlinear response is reflected in the fitting of the E-C relationship, which means that it cannot be performed with a function on a wide energy range. Linear fitting will be distorted at the low energy end (Zhang 2019). Therefore, piecewise fitting or polynomial fitting is generally used to deal with this problem. From the point of view of data analysis alone, this method is to select the fitting model based on only data. There is no physical meaning behind the fitting model. Moreover, the lower the energy end, the less accurate it is. Therefore, we need an explanation from the principle and mechanism to study the energy nonlinear response clearly and provide a fitting model with a physical basis. This will be very meaningful for the establishment of the calibration model. Only in this way can the data be accurately inverted and the response matrix can be more credible. On the other hand, due to the existence of nonlinear phenomena, higher requirements are put forward for the calibration test work. It is possible to obtain a more accurate energy response calibration of the detectors only under the enough and sufficient calibration points. Therefore, whether it is from the establishment of the calibration model or the consideration of the calibration experiment, the research on the nonlinear energy response of crystal will be very meaningful, and the current research on this aspect is very limited and not in-depth.

## 5.2 Data acquisition mode

During the entire calibration experiment, two data acquisition modes were used to collect experimental data, namely the normal mode (mode-I) and the dynamic baseline subtraction mode (mode-II). In this chapter, we will discuss the impact of different data acquisition modes on the energy response of the detector and the calibration of detection efficiency.

Select the data of Z01-29 for analysis. The data of 40-160 keV is collected by the data acquisition board in the normal mode, and the data of 38-70 keV and 80-160 keV are collected by the data acquisition in the dynamic baseline subtraction mode. Use the characteristic peak of 37.4 keV from background spectrum of the high gain channel to correct the gain of the high gain channel, and use the characteristic peak of 1470 keV from background spectrum of the low gain channel to correct the gain of the low gain channel. The comparison of E-C relationship, E-ER relationship, and detection efficiency of the data is shown in Figure 5.4 to 5.7. According to the analysis results, it can be seen that the two different data acquisition modes will not affect the calibration of the GRD energy response. The E-C relationship and the E-ER relationship are basically the same in the two modes. The measured values of detection efficiency in the two modes are matched with the simulated values. GRD energy response is not affected by the change of data acqui-sition mode.

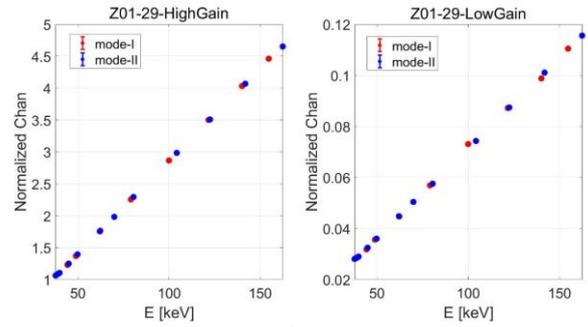

**Figure 5.4**   E-C relationship: HighGain (left) vs. LowGian (right), where mode-I is the normal mode and mode-II is the dynamic baseline deduction mode. In this figure, the normal mode data has been deducted the baseline in advance.

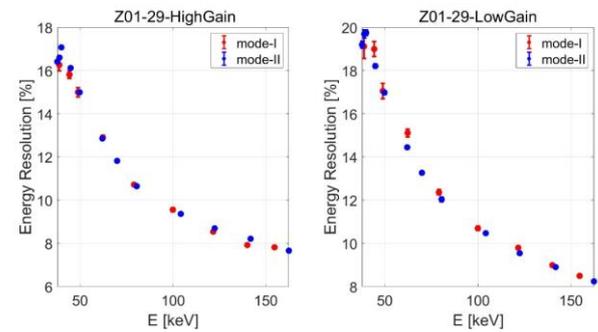

**Figure 5.5**   E-ER relationship: HighGian (left) vs. LowGian (right) (about 40 keV data dropped because there is an absorption edge in that energy).

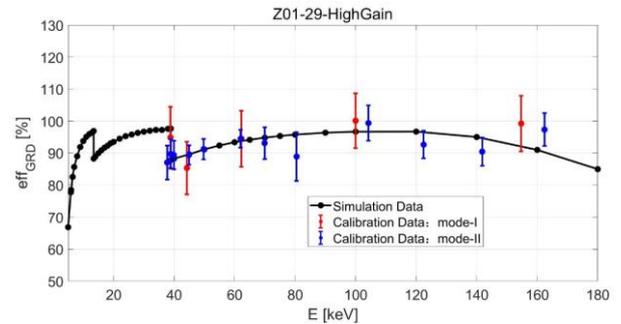

**Figure 5.6**   Comparison of detection efficiency in different modes (the black dotted line is the MC simulation value, the red dot is the data collected in the normal mode, and the blue dot is the data collected in the dynamic baseline subtraction mode): HighGain.

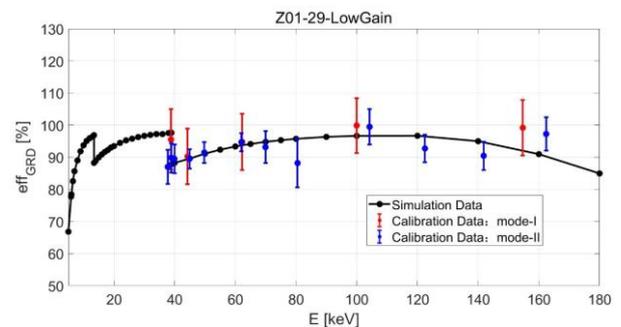

**Figure 5.7**   Comparison of detection efficiency in different modes (the black dotted line is the MC simulation value, the red dot is the data collected in the normal mode, and the blue dot is the data collected in the dynamic baseline subtraction mode): LowGain.

## 5.3   Working mode

The working mode of each GRD unit is divided into full-mode (Full-Mode), semi-mode 1 (Semi-Mode I) and semi-mode 2 (Semi-Mode II). The specific situation is shown in Table 5.1. The test dynamic range of the GRD corresponding to the full-component mode and the half-component mode is different, and it can be switched freely according to the needs of scientific testing. On-orbit GRD is powered on, it will work in the full-component mode, that is, supply power to all SiPMs of the GRD detector. This has a better detection effect for low-energy gamma-rays. Semi-component mode 1 and semi-component mode 2 are in some cases, if some SiPM is in short-circuit failure mode or need to expand the energy range measurement range, GRD can switch the SiPM power supply mode to one of two semi-component modes powered by half of SiPM for work. The GRD full-component mode is the main working mode that requires detailed calibration, and the two half-component modes also require simple calibration on the ground. The calibration method is to sample one official version GRD (Z01-61) to test the energy response and detection efficiency of the two half-components. The comparison test between the half-component and the full-component is carried out in the National Institute of Metrology, China (Changping). The test energy range is 8-40 keV. In this chapter, based on the E-C relationship, energy resolution, and detection efficiency in different modes, the influence of different operating modes on the energy response and detection efficiency of GRD will be discussed.

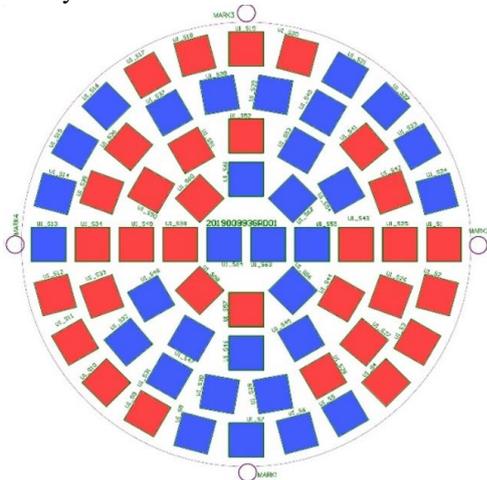

**Figure 5.8**   The distribution diagram of SiPM front amplifier circuit board (red: Semi-Mode I, blue: Semi-Mode II).

**Table 5.1**   GRD operating mode classification.

| | GRD Operating mode | Directions | Energy range |
|---|---|---|---|
| 1 | Full-Mode | All SiPM powered for working | 4 keV-4.5 MeV |
| 2 | Semi-Mode I | The first half of SiPM is powered for working | 8 keV-9 MeV |
| 3 | Semi-Mode II | The other half of SiPM is powered for working | 8 keV-9 MeV |

### E-C relationship

According to the ideal situation, the full-mode peak position value/2 should be equal to the value of the semi-mode under the same energy point. Taking the Z01-61 as the data sample, since the test energy point covers 8-40 keV, Therefore, this section only uses the data of the high-gain channel for discussion. Set the full energy peak value/2 to obtain the E-C relationship between the full-mode and the two semi-modes as shown in Figure 5.9 (the gain has been corrected by the characteristic peak from the background spectrum in advance).

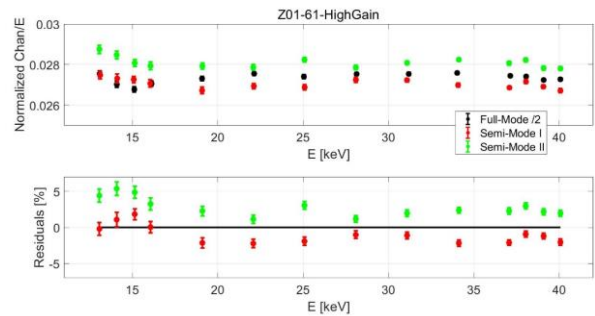

**Figure 5.9**   Comparison of E-C relationship in different working modes: Full-Mode/2 vs. Semi-Mode I vs. Semi-Model II, high gain channel.

According to the comparison results in Figure 5.9, the measured results under different working modes are basically consistent. Here Residual is defined as Residual = (Semi-Mode-Full-Mode)/ Full-Mode, the relative error of Semi-Mode I and Semi-Mode II with respect to Full-Mode/2 is within 5%, and the measured results can be considered to be consistent with theoretical expectations.

### Energy resolution

Under reasonable circumstances, the energy resolution Multiply by $\sqrt{2}$ in the full-mode working mode should be equal to the energy resolution in the semi-mode working mode. The comparison of energy resolution in the full-mode Multiply by $\sqrt{2}$ with the energy resolution in the semi-mode is shown in Figure 5.10.

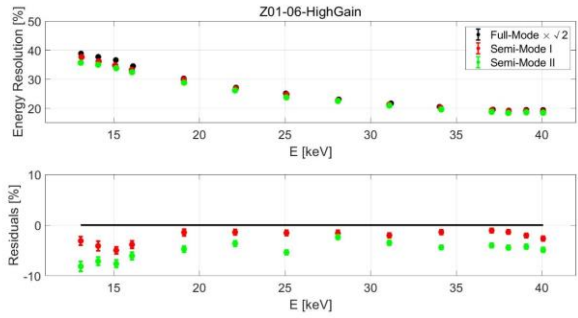

**Figure 5.10** Energy resolution comparison under different working modes: Full-Mode× $\sqrt{2}$ vs. Semi-Mode I vs. model2, high gain channel.

According to the comparison results in Figure 5.10, the measured results under the two semi-component working modes are basically consistent. The definition of Residual is the same as Figure 5.9. The relative change percentage of Semi-Mode I and Semi-Mode II with respect to Full-Mode× $\sqrt{2}$ is large below 15 keV. As the energy increases, the relative change value is within 5%, the actual measurement results are in line with theoretical expectations.

**Detection efficiency**

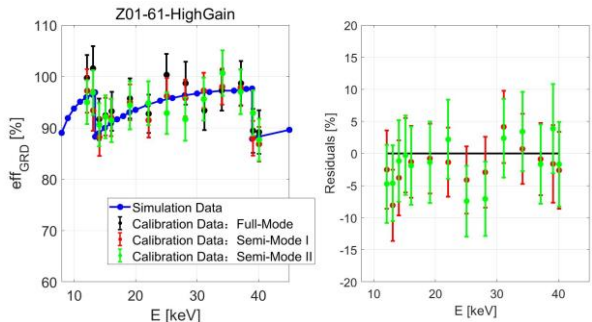

**Figure 5.11** Comparison of detection efficiency in different working modes: Full-Mode vs. Semi-Mode I vs. model2, high gain channel.

Figure 5.11 shows the measured values of detection efficiency under different working modes, and the Residual definition is the same as Figure 5.9. The measured values of the detection efficiency in the three modes all match the simulated values. The relative deviation of values in the two half-component modes and the values in the full-component mode are within 5%, which can be considered as the detection efficiency of GRD in different working modes is basically the same.

From the test results (Figure 5.9, 5.10, 5.11), the gains of semi-component mode 1 and semi-component mode 2 are very similar. From the peak point of view, the peak ratio of semi-mode 1, semi-mode 2 and full component mode is close to 1:2; in terms of energy resolution, the energy resolution ratio of semi-mode 1, semi-mode 2 and full component mode is close to 1: $\sqrt{2}$, and with the increase of energy, this ratio is more closer to 1: $\sqrt{2}$. In terms of detection effi-ciency, since only half of the SiPM works in the semi component, the relative threshold is relatively high. During the test, the semi component mode can be completely separated from the noise until to the energy of 13 keV, while under the full component mode, 9-10 keV can be separated from the noise, but the detection efficiency is consistent within the effective measurement range.

## 5.4 Spatial response

In the spatial response test of the GRD, in order to understand the uniformity of the x-ray response of each position of the detector, the GRD official version Z01-43 was tested for different position responses. There are two measurement energy points (15 keV and 39 keV) in response to the position of GRD, each energy point has a test time of 120 seconds, and the statistic is greater than 100,000 counts. The incidence of different positions (25 points in a rice-shaped distribution) is adopted. See the specific plan Figure 5.12. This test is carried out in the National Institute of Metrology, China. The parameter coordinates of each point are set, and the displacement platform is moved to make the light exit point to be tested respectively at the location to be tested. In addition, two semi-component mode related experiments were carried out at 19 keV energy.

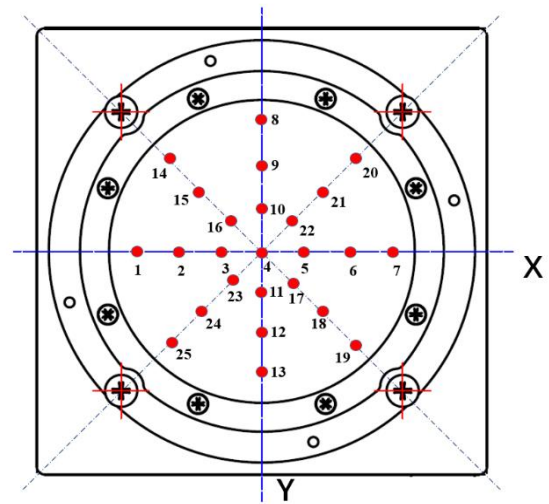

**Figure 5.12** Point distribution and location number definition schematic diagram of GRD spatial response test.

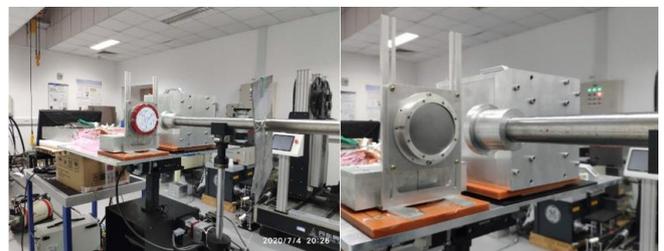

**Figure 5.13** Spatial response test on site.

Firstly, the full energy peak value at each energy and position can be obtained through the data analysis, the spectral broadening and area of full energy peak can be obtained by the fitting, then the uniformity of the GRD at different positions under the energy point can be obtained. Among them, the two energy points of 15 keV and 39 keV were tested in the full-component mode, and the energy point of 19 keV was tested in the two semi-component modes. The corresponding situation of test serial number and position is shown in Table 5.2.

**Table 5.2** Correspondence of test serial number and position (The numbers also represent the time sequence of the experiment).

| Number | 1 | 2 | 3 | 4 | 5 |
|---|---|---|---|---|---|
| Position | x-4 （0,0） | y-3 （0, -11） | y-2 （0, -11） | y-1 （0, -11） | m-3 （7.8, 25.2） |
| Number | 6 | 7 | 8 | 9 | 10 |
| Position | n-5 （0,15.6） | x-3(3.2,-7.8) | m-2 （4.6, -15.6） | n-6 （0,31.2） | x-2 （6.4,-15.6） |
| Number | 11 | 12 | 13 | 14 | 15 |
| Position | m-1 （1.4, -23.4） | x-1 （9.6,23.4 ） | n-7 （-9.6,23.4） | y-7 （-23.4,9.6） | y-6 （0,-11） |
| Number | 16 | 17 | 18 | 19 | 20 |
| Position | y-5 （0,-11） | y-4 （0,-11） | n-3 （-7.8, -7.8） | m-5 （0,15.6） | x-5 （-3.2,-7.8） |
| Number | 21 | 22 | 23 | 24 | 25 |
| Position | n-2 （-4.6,-15.6） | m-6 （0,31.2） | x-6 （-6.4,-15.6） | n-1 （-1.4, -23.4） | x-7 （-9.6,23.4） |
| Number | 26 | 27 | | | |
| Position | m-7 （9.6,23.4） | x-4 （+23.4,-23.4） | | | |

Analyze the energy spectra obtained at different positions to obtain the peak position and broadening. The relationship between the peak value and the position is shown in Figure 5.14. According to the results, 39 keV full-component mode peak value > 2 times 19keV semi-component mode peak value > 15 keV full-component mode peak value, semi-component mode 1 peak values are equal to semi-component mode 2 peak Values, consistent with theoretical expectations. And in the full-component and two semi-component modes, the peak values of each point of the GRD are basically the same, and the relative error is within 14% (3.6%@15 keV, 3.6%@39 keV, 13.8%@19keV semi-component mode 1, 11.5%@19 keV semi-component mode 2). Excluding the central energy point, the relative error is within 14% (1.5%@15keV, 1.4%@39 keV, 13.8%@19 keV semi-component mode 1, 11.5%@19 keV semi-component mode 2).

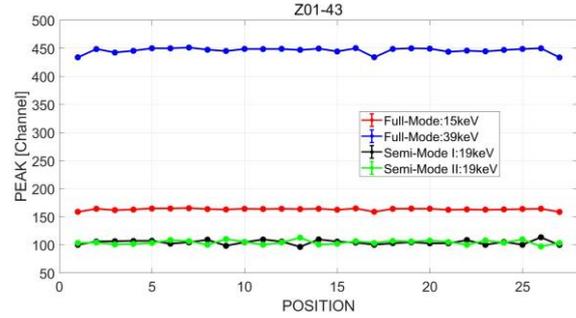

**Figure 5.14** Spatial response test results: peak value(red point is the result of 15keV in full-component mode, blue point is the result of 39 keV in full-component mode, black point is the result of semi-component mode I at 19 keV, green point is semi-composition mode II at 19 keV).

The relationship between the energy resolution and the position is shown in Figure 5.15. According to the results, 39 keV full-component mode energy resolution <19 keV semi-component mode energy resolution/ $\sqrt{2}$ <15 keV full-component mode energy resolution, semi-component mode 1 energy resolution = semi-component mode 2. The energy resolution is consistent with theoretical expectations. And in the full-component and two semi-component modes, the energy resolution of each point of the GRD is basically the same, and the relative error is within 9% (4.4%@15 keV, 8.4%@39 keV, 8.4%@19 keV semi-component Mode 1, 7.8%@19 keV semi-component mode 2). Excluding the central energy point, the relative error is within 8% (1.3%@15 keV, 1.0%@39 keV, 6.6%@19 keV semi-component mode 1, 7.8%@19 keV semi-component mode 2).

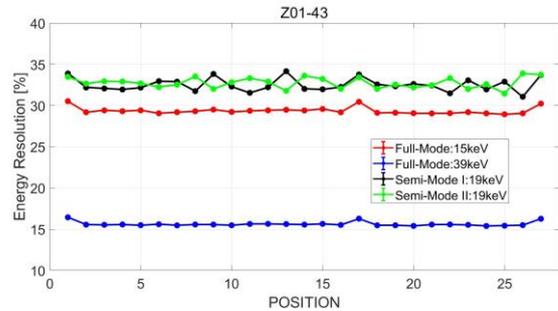

**Figure 5.15** Spatialresponse test results: energy resolution (red point is 15 keV result in full-component mode, blue point is 39 keV result in full-component mode, black point is the result in semi-component mode I at 19 keV, green point is the result in semi-component mode II at 19 keV).

Take the sum of counting rates in the range of 3σ around the full energy peak as the peak area for comparison. The relationship between the peak area and the position is shown in Figure 5.16. According to the results, the peak areas in the two semi-component modes are basically the same, and smaller than the peak areas in the full-component mode. The peak area in full-composition mode decreases with increasing energy. The peak areas of the three working modes have gradually become smaller with the pass of time. This result also verifies that the beam intensity of the X-ray beam is unstable, and with the pass of time, the beam inten-sity gradually weakened. At the same time, the rationality of the introduction of the $K_I$ factor in Chapter 3.3 of this paper is also verified.

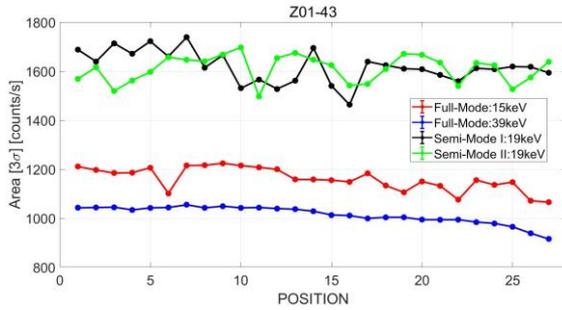

**Figure 5.16** Spatialresponse test results: peak area (red point is the result in full- component mode at 15 keV, blue point is the result in full-component mode at 39 keV, black point is the result of semi-component mode I at 19 keV, green point is the result of semi-composition mode II at 19 keV).

# 6 Conclusions

As the main detector of the GECAM satellite, the calibration of the energy response and detection efficiency of the GRD detector is the main content of the ground-based calibration. The calibration content requires the measurement energy points to cover 8 keV-2 MeV as much as possible. The statistics of energy points are more than 100,000 counts number. Individual GRDs are needed to accurately measured the absorption edges (with 0.1 keV as the step size). The low-energy band (8-160 keV) is calibrated with the X-ray beam in the National Institute of Metrology, China. The high-energy band (greater than 160 KeV) is calibrated with radioactive sources. In this article, we mainly focus on the calibration of energy response and detection efficiency in the 8-160 keV energy range.

During the calibration process, the ambient temperature is controlled at 20-25℃, and the temperature fluctuation is <±2℃. The focus is on the energy response and detection efficiency of the GRD under 8-160 keV X-ray beam. The main content has been discussed in Chapter 3: The E-C relationship, energy resolution and detection efficiency of 5 GRD samples are given. According to the analysis results in Chapter 3, the E-C relationship of the high gain/low gain channels of the GRD can be fitted with a quadratic function, and the specific model parameters have been given in the chapter. In this article, the relationship between energy resolution (ER) and energy (E) is fitted with a model of ER=P1+P2/ $\sqrt{}$ E. The measured detection efficiency of the GRD is consistent with the results given by the Monte Carlo simulation. Only around 160 keV, the measured value is slightly higher than the simulated value. We think this is related to the limitation of the efficiency curve of high-purity germanium.

Because lanthanum bromide crystals have absorption edges at 13-15 keV and 38-41 keV (Ivan V Khodyuk and Pieter Dorenbos 2010), the lanthanum bromide crystals around the absorption edges have been finerly tested. As shown in Chapter 4, using Z01-08 data as the sample, the energy response and detection efficiency of the high/low gain channels near the lanthanum bromide absorption edge are analyzed respectively. First, the quadratic function of the E-C relationship is fitted near the bromine absorption edge (13.48 keV) and the lanthanum absorption edge (38.89 keV) with a step length of 0.1 keV. This model is compared with other energy ranges (8-160 keV after removing the absorption edge), shows a more obvious difference. In the terms of the energy resolution, the phenomenon near the absorption edge is also significantly different from the performance of other energy bands. At the same time, according to Figure 4.7-4.8, it can be seen that the measured detection efficiency near the absorbing edge is in good agreement with the MC simulation result.

According to literature research, lanthanum bromide crystals have obvious energy nonlinearity less than 15 keV (Zhang 2019), so it is necessary to perform detailed calibration at these energies to study the nonlinearity of the low-energy end of the GRD detector. We have classified the 50 mounted GRDs according to the crystal doping, which are single-doped crystals, middle-doped crystals, and double-doped crystals. For different types of crystals, the energy nonlinearity at the low energy end is obvious different, see the results in section 5.1 for details. Single-doped crystals take Z01-29 as the sample, middle-doped crystals take Z01-35 as the sample, and double-doped crystals take Z01-70 as the sample for comparison. The energy resolution results are shown in Figure 5.2. The energy resolution of double-crystals is better than that of middle-doped crystals, while the energy resolution of middle-doped crystals is better than that of single-doped crystals. In addition, the actual detection efficiency of the three GRDs represented by the above three crystal types is consistent with the simulated value within the error range, which shows that the physical model given by Monte Carlo is credible.

During the ground-based calibration of the GRD detector, the data acquisition method adopts two modes, the normal mode and the dynamic baseline subtraction mode. Section 5.2 of this article discusses the impact of the data acquisition mode on the energy response and detection efficiency. Conclusion is, the data acquisition mode does not affect the E-C relationship, energy resolution, and detection efficiency.

The working mode of each GRD unit is divided into full-mode (Full-Mode), semi-mode 1 (Semi-Mode I) and semi-mode 2 (Semi-Mode II). When GECAM is running on orbit It can be switched freely according to the needs of scientific testing. Section 5.3 analyzes the influence of different working modes on the energy response and detection efficiency of the detector. The conclusion is (1) The E-C relationship under different working modes is in line with theoretical expectations. The relative change of Semi-Mode I and Semi-Mode II with respect to Full-Mode/2 is within 5%. (2) In the two semi-component working modes, the measured results of energy resolution are basically the same. The relative change percentage of Semi-Mode I and Semi-Mode II with respect to Full-Mode× √2 is become large below the energy of 15 keV. As the energy increases, the relative change value is within 5%. The measured result is consistent with theoretical expectations. (3) In the three working modes, the measured value of detection efficiency matches the simulated value within the error range, and the value in the two semi-component modes is within 5% of the value in the full-component mode, which can be considered that the detection efficiency under different working modes is basically the same.

In the spatial response test of the GRD, in order to understand the uniformity of the x-ray response of each position of the detector, Z01-43 was tested in different positions. In the full-composition mode, the deviation of the peak value between the different position points is less than 1.5% except for center point. In the semi-component mode, since SiPM works less than half, it is close to the lower energy threshold at 19 keV, and the maximum peak deviation is close to 14%. In terms of energy resolution, excluding the central point, the relative error is within 8%. In terms of relative detection efficiency (peak area), the maximum deviation of relative detection efficiency is close to 10.4% at 15keV in full-component mode, and the maximum deviation is close to 2.7% at 39 keV. When the semi-component mode is 19 keV, the maximum deviation of the relative detection efficiency between mode 1 and mode 2 is close to 12%. Excluding the instability of the system and other reasons, the overall counting rate uniformity of the GRD is relatively good, and the uniformity becomes significantly better after the energy is increased.

This article summarizes the ground-based calibration and special test of the detector GRD of GECAM satellite, which introduces the calibration results of the energy response and detection efficiency of the GRD ground-based calibration. The calibration results are reliable and meet the requirements of the GRD ground-based application system for calibration products. The calibration campaign is comprehensive and the measured data results are generally consistent with the simulations.